\documentclass[graybox]{svmult}

\usepackage{mathptmx}       
\usepackage{helvet}         
\usepackage{courier}        
\usepackage{type1cm}        
\usepackage{makeidx}         
\usepackage{graphicx}        
\usepackage[bottom]{footmisc}

\newcommand{\oversim}[2]{\protect{\mbox{\lower0.5ex\vbox{%
   \baselineskip=0pt\lineskip=0.2ex
   \ialign{$\mathsurround=0pt #1\hfil##\hfil$\crcr#2\crcr\sim\crcr}}}}} 
\newcommand{\simless} {\mbox{$\,\mathrel{\mathpalette\oversim<}\,$}} 
\begin{document}

\pagenumbering{arabic}

\title*{Lessons from the Local Group (and beyond) on dark matter}
\author{Pavel Kroupa}

\institute{Helmholtz-Institut f\"ur Kern und
  Strahlenphysik, University of Bonn, Nussallee 14--16, 
53115 Bonn, Germany; e-mail: pavel@astro.uni-bonn.de}
%
%
\maketitle


\abstract{The existence of exotic dark matter particles outside the
  standard model of particle physics constitutes a central hypothesis
  of the current standard model of cosmology (SMoC). Using a wide
  range of observational data I outline why this hypothesis cannot be
  correct for the real Universe. Assuming the SMoC to hold, (i) the
  two types of dwarf galaxies, the primordial dwarfs with dark matter
  and the tidal dwarf galaxies without dark matter, ought to present
  clear observational differences. But in fact there is no
  observational evidence for two separate families of dwarfs, neither
  in terms of their location relative to the baryonic Tully-Fisher
  relation nor in terms of their radius--mass relation. This result is
  illuminated by the arrangements of the satellite galaxies around
  host galaxies for which we have data: the arrangements in rotating
  disk-of-satellites, in particular around the Milky Way and
  Andromeda, has been found to be only consistent with most if not all
  dwarf satellite galaxies being tidal dwarf galaxies. The predicted
  large numbers of independently or in-group accreted,
  dark-matter-dominated primordial dwarfs are most inconspicuously
  absent around the Milky Way in particular. The highly symmetric
  structure of the entire Local Group too is inconsistent with its
  galaxies stemming from a stochastic merger-driven hierarchical
  buildup over cosmic time.  (ii) Dynamical friction on the expansive
  and massive dark matter halos is not evident in the data: the
  satellite galaxies of the Milky Way with proper motion measurements
  have no infall solutions as they would merge with the MW if they
  have dark matter halos, and galaxy groups such as the M81 group are
  found to not merge on the short time scales implied if each galaxy
  has a dark matter halo. {\it Taking the various lines of evidence
  together, the hypothesis that dynamically relevant exotic dark
  matter exists needs to be firmly rejected.}}

\pagestyle{plain}

\section{Introduction}
\label{sec:intro}

\vfill
\includegraphics{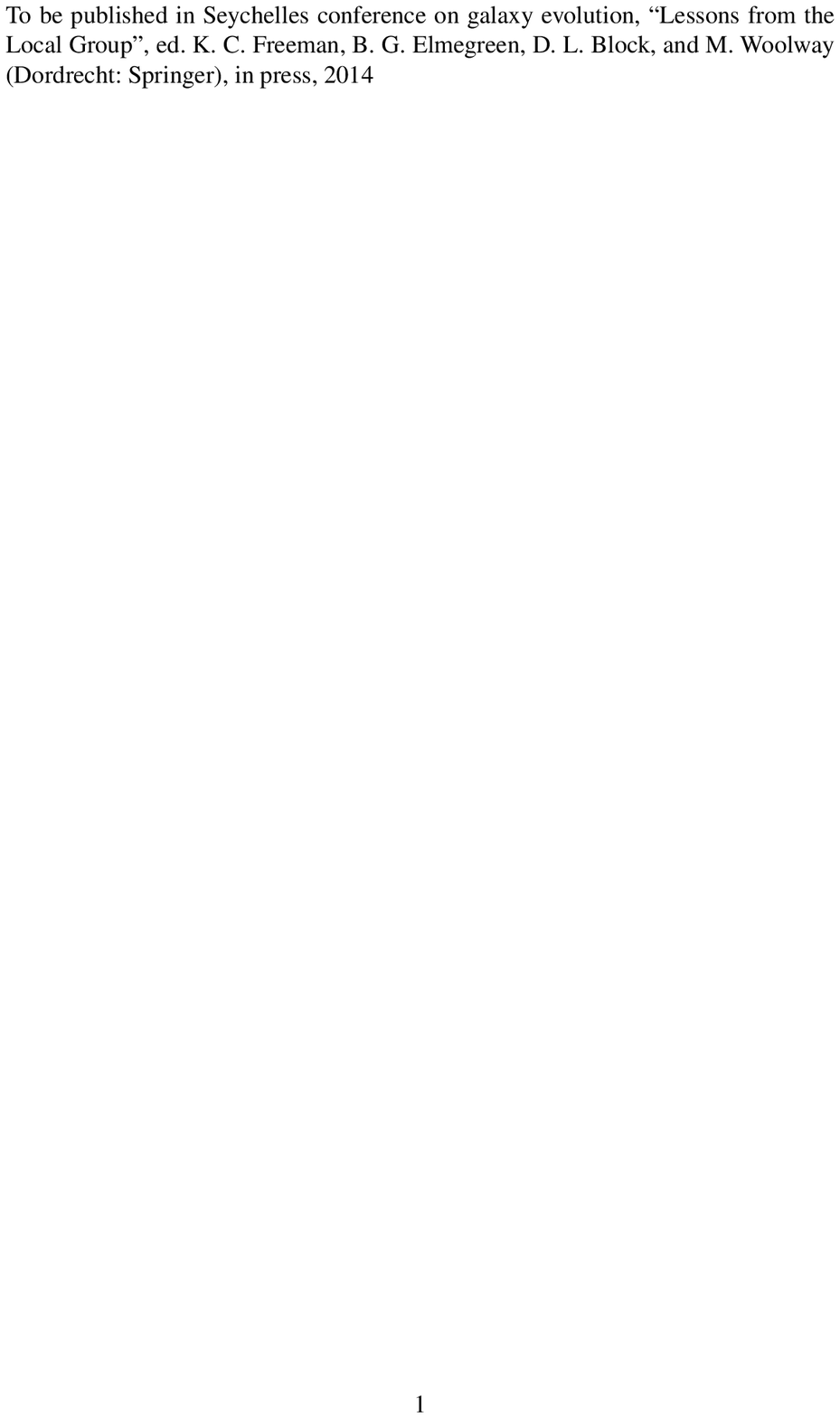}

By applying Einstein's general relativistic field equation,
i.e. Newtonian dynamics, to galaxies and to the Universe as a whole,
disagreements with observational data had been found that require the
additional assumptions of inflation, exotic dark matter particles and
dark energy. These may constitute major new physics components, but
none have supporting experimental evidence independently of the
observational astronomical data that are used to make the three
postulates. For example, dark matter particles are not contained in
the standard model of particle physics and have not been found in any
of the experiments designed to search for them. If they do not exist,
then a major pillar of the modern standard modell of cosmology (SMoC)
collapses, such that the SMoC would be ruled out as a representation
of the Universe and of the structures that develop within it over
cosmic time. ``{\it Does dark matter exist}?'' is thus one of the most
important questions of modern science. Direct searches for dark matter
particles cannot falsify this question by design, since a non-detection
may merely imply that the interaction cross section with baryons
(e.g. via the weak force) is unmeasurably small. Direct detection
experiments thus speculate on receiving the Nobel Price, but they are
not a well designed experimental procedure in which a prediction can
be falsified.

Here I argue that the astronomical observational data {\it strongly},
if not unequivocally, show {\it dark matter to not be present}. I use
three independent tests and many consistency checks. While this goes
against the perceived majority opinion with corresponding sociological
and possible career implications, the community does have to face a
reality without dark matter, as bleak and dark as it may appear.

Weighty evidence for this conclusion comes from the best data
at hand, namely what we learn from observing the galaxies and their
star-formation processes in the Local Group. But extragalactic
evidence has also been crucial in refining the conclusions.

\begin{figure}
\hspace{-6mm}
\includegraphics[width=13cm, angle=0]{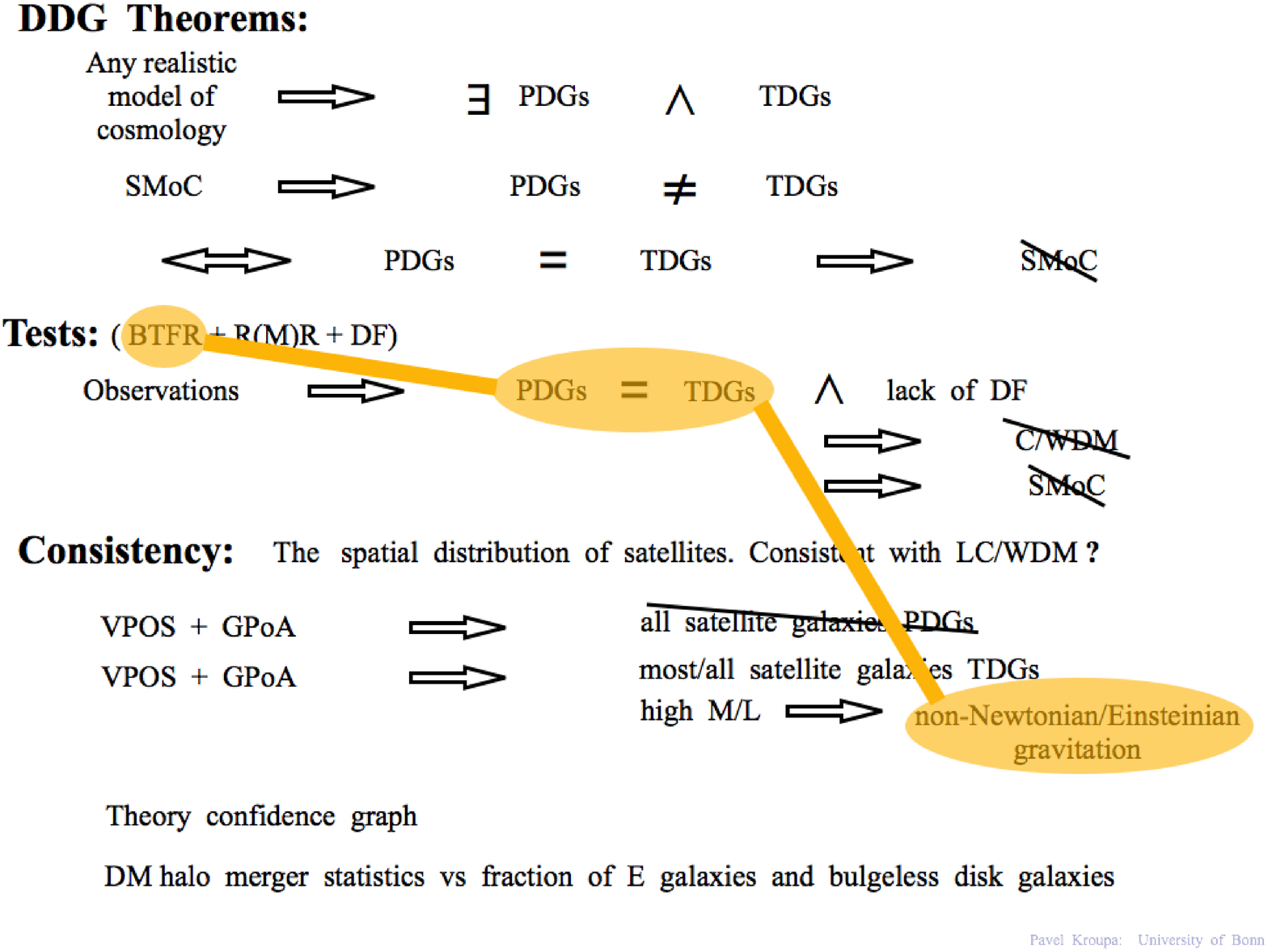}
\caption{\small Structure of the falsification of dynamically relevant
  cold or warm dark matter. The dual dwarf galaxy (DDG) theorems and
  the tests using the baryonic Tully Fisher relation (BTFR), the
  radius-mass relation (R(M)R) of pressure supported dwarf galaxies
  and using dynamical friction (DF) are discussed in
  Sec.~\ref{sec:noDM}.  The consistency checks based on the vast polar
  structure (VPOS) of the Milky Way and on the great plane of
  Andromeda (GPoA) are in Sec.~\ref{sec:consist1} (see also
  Fig.~\ref{fig:LG}). A high degree of self-consistency of the
  argument emerges by the VPOS and GPoA, which are mutually correlated
  rotating structures, ruling out the satellite galaxies being
  primordial dwarf galaxies (PDGs) with dark matter halos. Thus their
  observationally deduced high dynamical M/L ratios needs to be
  accounted for by effective non-Newtonian gravitation, which is the
  same conclusion reached using the BTFR test. This consistency is
  emphasized by the connected orange regions. If this conclusion is
  true then the SMoC cannot be a good description of the observed
  Universe.  The theory confidence graph for the SMoC is discussed in
  Sec.~\ref{sec:consist2} and indeed confirms the SMoC to not be an
  acceptable model; neither CDM nor WDM thus exists. This is further
  ascertained by the evidence for a lack of mergers in the observed
  galaxy population as covered in Sec.~\ref{sec:galpop}, whereby DM
  halo merger statistics are discussed in \cite{Kroupa14,WK14}.  Other
  acronyms: TDGs=tidal dwarf galaxies (Sec.~\ref{sec:noDM}),
  C/WDM=cold or warm dark matter= exotic DM particles, i.e. particles
  outside the standard model of particle physics. }
\label{fig:flogics}
\end{figure}

With this text I provide a synopsis of the arguments presented in more
depth in \cite{Kroupa12a} and \cite{Kroupa14}.  Fig.~\ref{fig:flogics}
outlines the structure of the argument.\footnote{Noteworthy is that
  the evidence provided in the year~2012 which are claimed to falsify
  the SMoC \cite{Kroupa12a} have, to this date, not been countered by
  the community but have instead been strengthened by recent progress
  \cite{Ibata14,Pawlowski14,Kroupa14}.  }

There are three main reasons why cold or warm dark matter (DM)
particles, collectively referred to here as exotic DM particles,
cannot be dynamically relevant on the scales of galaxies. These are
discussed in Sec.~\ref{sec:noDM}. Secs.~\ref{sec:consist1}
and~\ref{sec:consist2} contain consistency checks, and
Sec.~\ref{sec:galpop} develops additional arguments to test the
``no-DM'' deduction based on galaxy populations over cosmic time. The
conclusions are provided in Sec.~\ref{sec:concs}

\section{Why can there be no cold or warm dark matter?}
\label{sec:noDM}

Assume the SMoC applies, and view the Universe through Newtonian eyes.
By assuming the SMoC applies, we accept the cosmological merger tree
to be a description of how the DM halos and the galaxies within them
grow through many mergers due to dynamical friction on the DM
halos. The first structures begin to form before recombination and
many of them become primordial dwarf galaxies (PDGs). In the SMoC
these are contained in DM halos. Ignoring the well-known problem of
downsizing, namely that dwarf galaxies are observed to be younger but
ought to be older than more massive galaxies
\cite{Recchi09,Vogelsberger14,Vogelsberger14b}, the evolution of the
galaxy population can be followed \cite{Lu12,Vogelsberger14b}.  As the
larger DM halos build up through coalescence, the galaxies within them
merge. The gaseous and stellar matter that is expelled and which
carries away the angular momentum and energy from the merger in the
form of tidal arms often fragments and forms star clusters and new
dwarf galaxies. These dwarf galaxies are called tidal dwarf galaxies
(TDGs), and they may contain captured previously existing stars and
they may form from pre-enriched gas, but this depends on the
evolutionary stage of the pre-merger galaxies.

Differently to PDGs, TDGs do not contain significant amounts of DM,
because TDGs have such small masses ($\simless10^9\,M_\odot$) that DM
particles from the much more massive DM halos of the pre-merger
galaxies are rarely captured by them. Thus, in any realistic
cosmological theory in which galaxies can interact, TDGs and PDGs must
exist. This is the ``{\it dual dwarf galaxy theorem}''
\cite{Kroupa12a,Kroupa14}. In the SMoC, TDGs and PDGs differ in their
DM content, and thus can be distinguished observationally.

Also, in the SMoC encounters between galaxies lead to them merging,
and therefore to the emergence of the cosmological merger tree which
drives galaxy evolution. Without dark matter halos, the merger rate
decreases most significantly \cite{Toomre77,Kroupa14}, and there would
be no cosmological merger tree which drives galaxy evolution. Our
understanding of galaxy evolution is thus intimately connected to the
existence of DM.

There are three tests which, independently of each other, rule out DM
as a relevant physical aspect of galaxies, as long as the
observational data remain undisputed. Two tests are based on the dual
dwarf galaxy theorem, and one is based on the well-understood process
of dynamical friction of the motions of galaxies through the DM halos
of other galaxies.

\begin{enumerate}

\item Tidal dwarf galaxies cannot contain much DM and yet the three
  that have observed rotation curves show the same DM behavior as
  PDGs. Thus, the three TDGs lie on the observed baryonic Tully-Fisher
  relation (BTFR) but they should be displaced by a factor of~3 to~10
  towards smaller velocities than the BTFR. Chance superposition of a
  TDG onto the DM-defined BTFR may occur if the velocity field of the
  gas in or around the TDG does not constitute a virialised
  (Keplerian) structure, but in such a case the TDG woud be more
  likely placed elsewhere in the BTF diagramme.  The rotational
  velocities of PDGs are assumed (but not understood) to be defined by
  the DM halos that are ten to hundred times more massive than the
  baryonic mass of the PDG \cite{Lu12,Vogelsberger14b}. That the
  baryonic masses do not correlate exactly one-to-one with the DM
  halos masses \cite{Behroozi13,Ferrero12} comes from the stochastic
  and haphazard process of the mergers which build-up the DM halo, and
  the different modes of accretion (cold vs hot) in DM halos of
  different masses.

  Since TDGs and PDGs lie on the same BTFR, which is supposed to be
  defined by primordial galaxies that are DM dominated, it follows
  that DM is ruled out to be the origin for the BTFR. This deduction
  is sound, as long as the data remain unchallenged, because TDGs
  cannot contain much DM, even if it exists, such that they cannot lie
  on the BTFR. But their rotation velocities can be obtained in a
  non-Newtonian gravitational framework (e.g. in Milgromian dynamics
  \cite{Gentile07}). PDGs can have DM, but their rotation curves can
  also be explained by non-Newtonian gravity. {\it Thus, non-Newtonian
    gravitation is the only unifying concept concerning TDGs and
    PDGs.}

\item Tidal dwarf galaxies must have different radii at the same mass
  than PDGs because they form without a DM halo, compared to
  pressure-supported PDGs which form within a substantial DM halo
  \cite{Behroozi13,Ferrero12}. But TDGs are found to have, at a given
  mass, the same radii as PDGs \cite{Dab13}. Since TDGs cannot contain
  DM this implies that the morphological appearance needs to be driven
  by a physical process common to both, the TDGs and PDGs. {\it Thus, only
  non-Newtonian gravitation can unify both types of dwarf galaxy.}

\item Dynamical friction is a necessary and required property of DM
  halos \cite{Toomre77,Barnes98}. Two similar galaxies that interact
  with relative velocities smaller than about the virial velocity
  dispersion of their DM halos and within a distance twice the virial
  radius of their DM halos will merge within an orbital time scale. A
  primordial satellite galaxy will merge with the main galaxy within a
  timescale given by Chandrasekhar's friction time scale.

  The dynamical simulations of the M81 group of galaxies has shown
  them to merge within a group-crossing time such that the matter
  bridges that are observed between the galaxies cannot be reproduced
  \cite{Thomson99,Yun99}. Reproduction by models of the observed
  bridges and galaxy locations and line of sight velocities is only
  approximately successful in models without DM. Since galaxies do
  have flat rotation cuves, it follows that these need to be explained
  without dark matter, i.e. {\it with non-Newtonian dynamics}.

  And, the satellite galaxies of the Galaxy with proper motion
  measurements cannot be traced back to pre-infall dwarf galaxies if
  they have DM halos \cite{Angus11}.

\end{enumerate}

\section{Consistency of the deduction: The arrangement of galaxies in
  the Local Group and elsewhere}
\label{sec:consist1}

If the deduction reached in Sec.~\ref{sec:noDM} that DM does not exist
is correct such that the cosmological merger tree would need to be
discarded, then what is the origin of the satellite galaxies that are
observed around the Milky Way (MW), Andromeda and other nearby
galaxies, and where do the dwarf galaxies, e.g. in the Local Group,
stem from? The lack of evidence for the cosmological merger tree being
active is also discussed in Sec.~\ref{sec:galpop}.

\begin{figure}
\hspace{-6mm}
\includegraphics[width=15cm, angle=0]{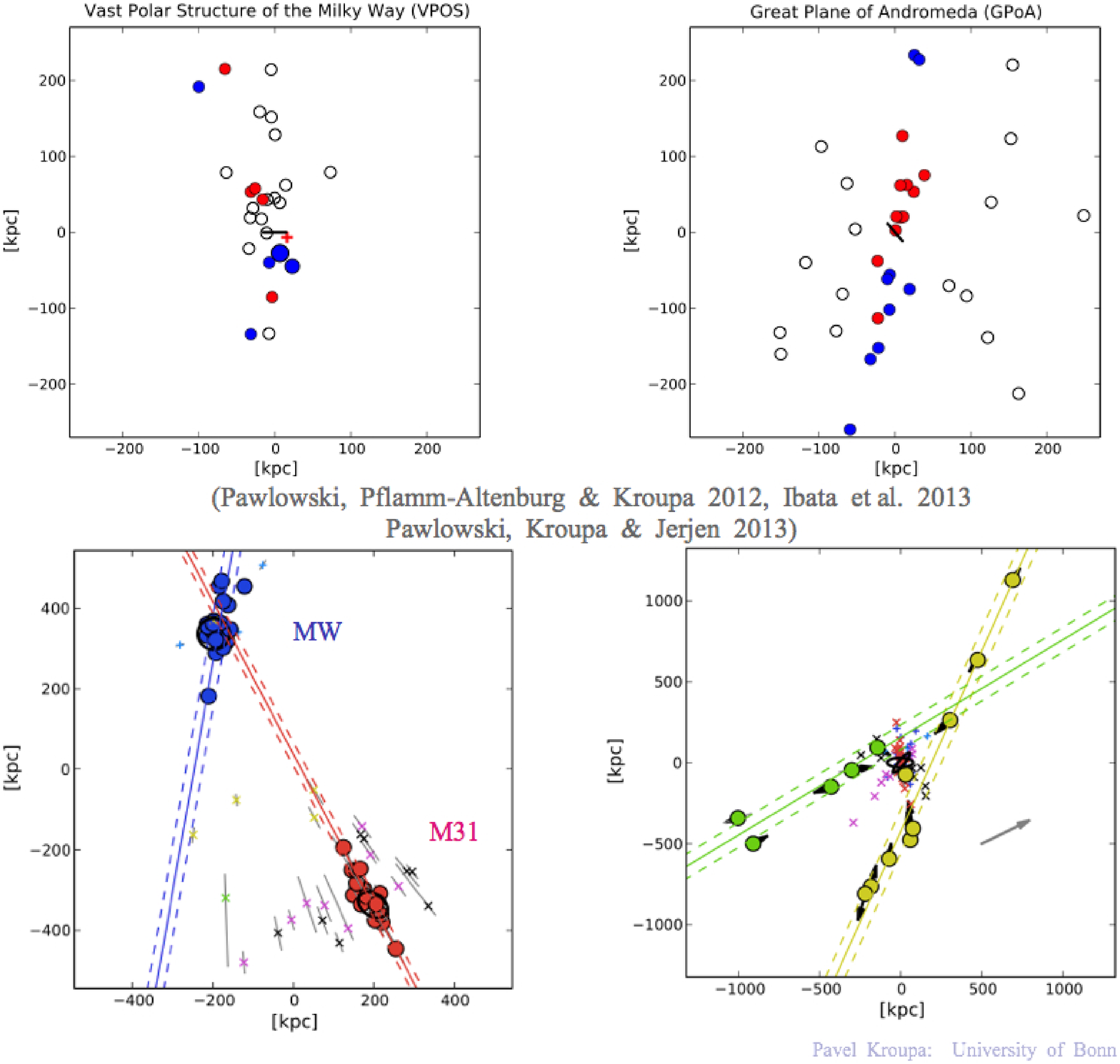}
\caption{\small The distribution of all galaxies known today in the
  Local Group which is defined by the zero velocity sphere.  Galaxies
  outside this sphere of radius about 1.5~Mpc recede from the Local
  Group, while within the sphere the galaxies fall towards us. {\it
    The upper left} and {\it right panels} depict all satellite
  galaxies within about 250~kpc around the MW and Andromeda,
  respectively. The galactic disks of the MW and of Andromeda are seen
  nearly edge-on, the north Galactic pole direction being upwards, and
  both galaxies are viewed from the same direction from infinity. MW
  satellites with known proper motions and radial velocities
  \cite{Pawlowski12a,Pawlowski13}, and Andromeda satellites that are
  in the GPoA \cite{Ibata13} are shown as colored circles. Red
  satellites are moving away from the observer, blue ones are
  approaching the observer. Thus the VPOS and GPoA are rotating in the
  same sense and the GPoA is seen edge-on from the MW as is evident in
  the lower-left panel. The VPOS and GPoA are statistically highly
  significant mutually correlated rotating structures inconsistent
  with being derived from accreted DM sub-halos which host PDGs
  \cite{Ibata14,Pawlowski14}. The VPOS and GPoA are inclined relative
  to each other by only about 38~degrees whereby the GPoA is oriented
  edge-on to the MW. This is seen in the {\it lower left panel}, which
  is a view from near the north Galactic pole downwards such that the
  VPOS and GPoA are seen approximately edge-on.  Filled circles are
  dwarf galaxies in the disks of satellites of both major galaxies,
  while the crosses near M31 are satellite galaxies which are not in
  the GPoA.  Viewing the Local Group along the line joining the MW
  and Andromeda it emerges that all non-satellite galaxies in the
  Local Group are arranged in two highly symmetrical equally thin
  planes (thickness about 50~kpc; diameters about 3~Mpc) equidistant
  from both, the MW and Andromeda, seen here edge on in the {\it lower
    right panel}. The crosses are as in the lower left panel; note the
  additional disk of satellite which lies in the equatorial plane of
  Andromeda and is here seen as a string of dwarfs (crosses) between
  the two major planes of the Local Group.  The green and yellow
  filled circles show all non-satellite dwarf galaxies comprising the
  Local Group. They are all situated in one of two major symmetric
  planes.  The short arrows depict the galaxy motions as given by the
  line-of-sight velocities.  The Local Group is moving along the arrow
  towards the CMB.  {\it Evidently the Local Group is a highly
    structured symmetric distribution of galaxies which has never been
    predicted nor even hinted at as being a result of the structures
    forming through a SMoC merger tree.}  The upper two panels are
  according to \cite{Kroupa14}, while the two lower panels are
  according to \cite{Pawlowski13b}.  }
\label{fig:LG}
\end{figure}

\subsection{Clues on their origin: spatially anisotropic satellite galaxy populations}

The spacial arrangement of satellite galaxies around the MW gives a
strong clue to their possible origin. {\it All} known stellar systems
beyond about 10~kpc distance from the Galactic centre (classical dSph
satellite galaxies found on photographic plates, ultra faint dwarf
galaxies found with robots scanning the sky, globular clusters) and
halve of all gas and stellar streams are arranged independently of
each other in a vast polar structure (VPOS,
\cite{KTB05,Pawlowski12a,Pawlowski14,Pawlowski14b}). The proper motion
measurements for the~11 brightest satellite galaxies show the VPOS to
be rotating in one sense, the spin of the VPOS points into a direction
which is close to the MW disk plane (lets call it direction
$\mathbf{S}$).  Viewing along the direction Galactic-centre---Sun, the
VPOS lies approximately face-on.  The VPOS can be described as a polar
disk with diameter of about 500~kpc and thickness of about 50~kpc.

Halve of all satellite galaxies of Andromeda are in a vast thin disk
of satellites (VTDS), i.e. in the great plane of Andromeda (GPoA,
\cite{Ibata13,Ibata14}). This GPoA is rotating, and its spin is
directed only about 38~deg away from $\mathbf{S}$. That is, the GPoA
and the VPOS are impressively aligned, with the GPoA also being nearly
perpendicular to the MW disk as is the VPOS (fig.~16 in
\cite{Pawlowski13b}).  Fig.~\ref{fig:LG} shows the VPOS, the Andromeda
satellite galaxy system and their relative orientation and location,
and {\it the most remarkable and hitherto not noticed nor ever
  expected symmetric structure of the entire Local Group.}

\subsection{Other dwarf satellite galaxy populations}

Beyond the Local Group, \cite{Chiboucas13} discuss the dwarf galaxy
population in the M81~group of galaxies, which is a sparse group
comparable to the Local Group, and they find evidence that the faint
satellite galaxies are distributed anisotropically. They write ``In
review, in the few instances around nearby major galaxies where we
have information, in every case there is evidence that gas poor
companions lie in flattened distributions.'' \cite{Kroupa14} counts
nine major galaxies with associated satellite systems which are
anisotropic (see also \cite{Pawlowski14b}). {\it It thus seems to be
  more the rule than the exception that satellite galaxies appear to
  be highly correlated in phase-space such that they appear arranged
  highly anisotropically about their hosts}.

\subsection{Satellite galaxies are tidal dwarf galaxies (TDGs)}

How can the preponderance of such highly correlated satellite galaxies
be explained? The occurrence of an anisotropic system of PDG
satellites which is comparable to the VPOS or the GPoA is so unlikely
that this possibility can be discarded safely, even if accretion of
PDGs from cosmological filaments is considered
\cite{Pawlowski12b,Ibata14,Pawlowski14}. The only known viable
physical process which can generate such correlated structures is if
the satellite galaxies are TDGs which formed in tidal arms produced in
a galaxy--galaxy encounter together with associated massive star
clusters. Computer simulations of galaxy encounters show such
structures to emerge readily
(e.g. \cite{BH92,Wetzstein07,Bournaud08,Bournaud10,Pawlowski11,Hammer13,Yang14}),
as shown by the pioneering work of Bruce Elmegreen et
al. \cite{Elmegreen93}.

The symmetric structure of the Local Group, and the mutually
correlated disks of satellites around both, the MW and Andromeda, may
have been created when the much younger Galaxy and Andromeda
interacted closely ($<$55~kpc) about 7--11~Gyr ago \cite{zhao13}.
This encounter would have thrown out gas rich tidal arms in which the
dwarf galaxies of the Local Group formed. It would have thickened the
then existing disk of the MW and of Andromeda and it may have led to
the rapid buildup of a MW pseudo bulge through an induced radial-orbit
instability (\cite{zhao13,Kroupa14}, see also the MSc thesis at
Cambridge university: \cite{Banik14}).

Can TDGs, once formed, survive for a Hubble time? Yes they can. This
has been shown by simulations that include star formation and feedback
by \cite{Recchi07} and \cite{Ploeckinger14}: self-consistent and thus
self-regulated TDG formation implies that they do not blow themselves
apart due to a star burst. TDGs may be destroyed on an orbital time
scale if they are on highly plunging orbits.  TDGs which have consumed
their gas or have been partially stripped off it also do not dissolve
easily. The simulations of gas free TDGs evolved dynamically over a
Hubble time of tidal stressing have demonstrated them to survive
\cite{Kroupa97,Casas12}. Such satellite galaxies loose most of their
stars but evolve into quasi-stable remnants which feign domination by
DM although they do not contain any, as has been discovered by
\cite{Kroupa97}. In that paper \cite{Kroupa97} a prediction of a
satellite galaxy was made which was discovered ten years later and is
today known as the Hercules satellite galaxy. The predicted model
agrees with the radius, the velocity dispersion, the luminosity and
dynamical $M/L$ ratio nearly exactly with the real Hercules satellite
galaxy (fig.~6 in \cite{Kroupa10}).  And, observations have led to the
discovery of a few~Gyr old TDGs \cite{Duc14}. 

{\it Thus, many and perhaps a majority of TDGs apear to survive over
  many Gyr such that the MW and Andromeda dSph and ultra-faint dwarf
  satellite galaxies may be ancient TDGs.}

\subsection{If they are TDGs, then there is no cold or warm DM}

The above then implies the following for the existence of exotic dark
matter, independently of the arguments made in Sec.~\ref{sec:noDM}:
Since the satellite galaxies that are in the GPoA have the same
morphological properties and the same high dynamical mass to light
($M/L$) ratios as the Andromeda satellite galaxies not in the GPoA,
the former and latter must be described by the same dynamics. If the
former are ancient TDGs then they cannot contain dark matter. Thus,
the ``dark matter effect'', i.e. the elevated $M/L$ ratios, {\it can
  only be due to non-Newtonian dynamics}. Also, since the MW
satellites are all in the VPOS and because they have high $M/L$
ratios, again non-Newtonian dynamics needs to be invoked. That
Milgromian dynamics \cite{Milgrom83,FM12} accounts well for the
properties of the Andromeda and the MW satellites has been shown by
\cite{MW10,MM13,Lueghausen14}. Tidal modulation and shaping of TDGs as
they evolve on eccentric orbits over many Gyr additionally changes the
stellar phase space distribution function \cite{Kroupa97,Casas12} such
that even in Milgromian dynamics elevated apparent (but not true)
dynamical $M/L$ ratios are expected to result which deviate from the
pure-Milgromian values.\footnote{Although \cite{Kroupa97} suggested
  that the high $M/L$ ratios of the dSph satellite galaxies may be due
  to repeated tidal shaping of the stellar phase-space velocity
  distribution function in a Newtonian universe, this process is
  unlikely to account for all dSph satellite galaxies because they are
  on very different orbits. Consequently, non-Newtonian dynamics is
  required to account for the observed dynamical masses of all dwarf
  satellites.}

Here the beautiful work by David Block et al. \cite{Block06} becomes
relevant as evidence: Block et al. have shown that the about~1~kpc
sized dust ring near the centre of Andromeda may be explained by the
compact and massive satellite M32 punching through the disk of
Andromeda about 210~Myr ago. Given the present-day stellar mass of M32
($\approx3\times10^9\,M_\odot$), it would have had a pre-infall DM
halo mass $>10^{11}\,M_\odot$ \cite{Behroozi13,Ferrero12}, such that
dynamical friction on the massive DM halo of Andromeda would have
significantly altered the orbit of M32. This problem thus needs to be
studied further.

\section{Consistency of the deduction: The performance of the SMoC and
the theory confidence graph}
\label{sec:consist2}

If there is no dark matter, then the SMoC cannot be a realistic
description of the Universe. How does the track-record of the SMoC in
accounting for observational data fare? If it were to be good, i.e.,
if there is a long history of predictions which have been verified by
observations performed after the prediction was published, then this
would contradict the conclusion reached above that challenges the
existence of exotic cold or warm dark matter. This has been studied
using the {\it theory confidence graph} \cite{Kroupa12a}. It turns out
that the SMoC has a long history of failed predictions. If each
failure or problem is associated with a reduction in confidence by
50~per cent in the fundamental theory (that Einstein's general
relativity is valid everywhere, and that all matter emerged at the big
bang), then the SMoC would currently retain a probability of being a
valid representation of the Universe of less than $10^{-5}$~per cent
\cite{Kroupa14}. This probability is further reduced taking into
account the additionally failed predictions since 2012, such as the
large-scale observational evidence against the cosmological principle
\cite{Clowes13}, and the observed significant under-density of matter
within the local volume of about 400~Mpc (fig.~7 in \cite{Kroupa14}).
Thus, the present-day very low confidence in the SMoC is in agreement
with the above no-DM conclusions. {\it Consequently the SMoC is not a
  physical representation of the real Universe.}

\section{The galaxy population}
\label{sec:galpop}

If there is no dark matter, then the SMoC does not describe the
Universe. The observed ``dark matter effects'' in galaxies then need
to be explained by an effective non-Einsteinian/non-Newtonian theory
of gravitation. In this case dynamical friction on DM would not occur
and mergers would be much rarer despite galaxy--galaxy
interactions. Can this be seen in the observed evolution of the galaxy
population? A few important results:

\begin{itemize}

\item \cite{Lu12} construct semi analytically computed populations of
  galaxies based on a SMoC merger tree and star-formation recipes
  trimmed to agree with broad observed properties by discarding
  parameter ranges. The best final trimmed model leads to a curved
  BTFR in disagreement with the observed BTFR and to a much larger
  number of satellite galaxies than is observed, among other
  problems. The milestone Illustris project, which is the currently
  highest existing resolution calculation of structure formation of
  the Universe and includes gas dynamics and detailed star formation
  prescriptions \cite{Vogelsberger14b}, yields a Tully-Fisher relation
  in disagreement with the one obtained by \cite{Lu12}. But this is
  not discussed nor is the disagreement clarified by
  \cite{Vogelsberger14b}.  Both are steeper than the observed relation
  for galaxies with stellar masses larger than about
  $10^{10}\,M_\odot$. That is, the model galaxies have larger
  rotational speeds at a given mass than the observed ones. An
  unphysical aspect of such models is that they require stellar
  feedback to be a function of the hosting DM halo in order to have
  sufficient feedback energy to stop a sufficient amount of baryons
  making stars immediately such that they can be blown out and
  re-accreted slowly thereby helping to build-up galactic disks.
  \footnote{This would imply, essentially, that the table in my dining
    room would know it exists in the MW DM halo rather than in the DM
    halo of the Large Magellanic Cloud, in violation of the required
    fundamental property of DM particles which are supposed to not
    interact, apart maybe weakly, with the particles of the standard
    model of particle physics.}

  In contrast, scale-invariant or Milgromian dynamics yields the
  observed Tully-Fisher relation exactly \cite{FM12,Kroupa14}.

\item \cite{Shankar14} perform semi-analytical modelling of early-type
  galaxy formation. An interesting result from this work is that they
  need to suppress dynamical friction for improved agreement with the
  observational data. This is consistent with the independently
  deduced absence of evidence for dynamical friction noted in
  Sec.~\ref{sec:noDM}, Test~3.

\item \cite{Weinzirl09} and \cite{Kormendy10} find the fraction of
  disk galaxies with classical bulges to be very small. The small
  fraction (6~per cent) of disk galaxies with classical bulges is
  supported by \cite{Fernandez14} for a sample of 189 isolated
  galaxies. \cite{Kormendy10} point out that the small fraction of
  disk galaxies with classical bulges is incompatible with the merging
  history which would affect most galaxies if the SMoC were true. One
  deduction from this would be that mergers therefore cannot be a
  major aspect of galaxy evolution. The only way to suppress the
  occurrence of galactic mergers is to discard the massive DM halos
  made of particles.

\item The population of galaxies is {\it vastly} dominated by
  late-type galaxies. According to \cite{Delgado10} only 3--4~per cent
  of all galaxies more massive in stars than about $10^{10}\,M_\odot$
  are elliptical. This holds for the galaxy population about 6~Gyr ago
  and at the present epoch, and is in excellent agreement with the
  long-known result that disk galaxies are the by far dominant
  population in the field as well as in galaxy clusters (see fig.~4.14
  in \cite{BM98}).  It has never been successfully demonstrated that
  the merger-driven buildup of the galaxy population in the SMoC leads
  to the observed massive preponderance of rotationally supported,
  thin-disk star-forming late-type galaxies. Instead, galaxies that
  form in the SMoC are predominantly of early type, because angular
  momentum is ejected or cancelled-out during the many mergers.  As
  emphasized by \cite{Disney08}, the vast majority of galaxies appear
  to be a one-parameter family of objects, much simpler than expected
  with little variation (see also \cite{Kroupa14}). Consequently,
  dark-matter-driven mergers cannot be a physically relevant process
  in galaxy formation.

\item That this dominating population of late-type disk star-forming
  galaxies lie on a main sequence, is discussed by
  \cite{Speagle14}. These authors show that the main sequence of
  galaxies which corresponds to an approximately constant specific
  star-formation efficiency (star-formation rate per unit stellar
  mass) has a small dispersion and persists to high redshift. Galaxies
  are thus much simpler than expected from the haphazard buildup
  through a DM-driven merger tree in the SMoC, a view already arrived
  at by \cite{Disney08} in their principle-component analysis of a
  large sample of galaxies.

\end{itemize}

The overall implication of this discussion is thus consistent with the
above conclusions that DM-driven processes do not appear to play a
role in the astrophysics of galaxies.

\section{Conclusions}
\label{sec:concs}

With exotic DM particles being ruled out by observation as being an
important aspect of galactic dynamics, there would be no reason to
consider the existence of such particles at all. Therewith the central
pillar of the SMoC collapses, and the SMoC becomes irrelevant as a
theoretical framework for the Universe.  Probably most aspects of
current cosmological understanding then collapse as well: the standard
redshift--age and redshift--distance relations would probably be
wrong, the inferred cosmological evolution of the star-formation rate
density and of galaxy masses and of their ages would probably be wrong
as well. 

The failures of the SMoC thus require a new paradigm which comes by
without exotic DM particles \cite{Kroupa12b}.  Important and
successful hints have become available through the famous work of
Milgrom \cite{Milgrom83,Milgrom09}. The possible connections between
the observed non-Newtonian but scale-invariant dynamics in the
weak-field limit and cosmological parameters and the physics of the
vacuum noted by Milgrom \cite{Milgrom99} may indicate a deep
interrelation of both. Based on such ideas, a conservative
cosmological model without exotic DM particles as described by
\cite{Angus09,AD11,Angus13} may be emerging \cite{Kroupa14}. As of
very recently and thanks to special funding from the rectorate of the
University of Bonn we have now, for the very first time, an adaptive
mesh-refinement code, {\it Phantom of Ramses} (POR), which includes
full treatment of baryonic processes \cite{Lueghausen14b}. With POR
cosmological structure formation simulations of a universe consisting
only of the constituents of the standard model of particle physics and
with Milgromian dynamics have become possible.

Irrespective of which cosmological model may be the next standard one,
it will have to account for the time-dependent distribution and motion
of matter on large-scales and on galaxy scales as well as for all
properties of the microwave cosmic background.

\section{Acknowledgments}

I thank Ken Freeman for organizing this conference on the Seychelles.
It will remain memorable for decades to come. I also thank David Block
and Bruce Elmegreen for being around so actively and for so many
years such that we could have this splendidly luxurious meeting to
honor both of them.

\bibliographystyle{unsrt}
\bibliography{/Users/pavel/PAPERS/MOND_CanadianJournalPhyscs/Submit_new/PAPER/kroupa_ref}

\end{document}